\documentstyle[preprint,aps]{revtex}

\def\be{\begin{equation}}
\def\ee{\end{equation}}
\def\bea{\begin{eqnarray}}
\def\eea{\end{eqnarray}}

\begin{document}
\title{HEAT CAPACITY MEASUREMENTS IN PULSED MAGNETIC FIELDS}
\author{M. Jaime$^1$, R. Movshovich$^1$, J. L. Sarrao$^1$, J. Kim$^2$, G. Stewart$^2$%
, W. P. Beyermann$^3$, and P. C. Canfield$^4$}
\address{$^1$Los Alamos National Laboratory, Los Alamos, NM 87545, USA}
\address{$^2$University of Florida, Gainesville, FL 32611-8440, USA}
\address{$^3$University of California, Riverside, CA 92521-0413, USA}
\address{$^4$Iowa State University, Ames, IA 50011, USA}
\date{World Scientific, to be published}
\maketitle

\begin{abstract}
The new NHMFL 60T quasi-continuous magnet produces a flat-top field for a
period of 100 ms at 60 Tesla, and for longer time at lower fields, e.g. 0.5
s at 35 Tesla. We have developed for the first time the capability to
measure heat capacity at very high magnetic fields in the NHMFL 60T
quasi-continuous magnet at LANL, using a probe built out of various plastic
materials. The field plateau allows us to utilize a heat-pulse method to
obtain heat capacity data. Proof-of-principle heat capacity experiments were
performed on a variety of correlated electron systems. Both magnet
performance characteristics and physical properties of various materials
studied hold out a promise of wide application of this new tool.
\end{abstract}

\tightenlines

\newpage%

\section{Technique}

The 60 Tesla Long-Pulse (60TLP) magnet was recently commissioned at the Los
Alamos National Laboratory. This magnet produces a flat-top field for a
period of 100 ms at 60 Tesla, and for longer time at lower fields, e.g. 0.5
s at 35 Tesla. During the entire pulse, the magnetic field varies at a
maximum ramp rate of $dB/dt\approx 400$ Tesla/sec. Together, these
properties allow for the development of brand new tools to study materials
in pulsed magnetic fields. Heat capacity measurement is one of these tools.
We have built a probe made of plastic materials that allows us to perform
heat capacity measurements at temperatures between 1.6 K and 20 K in fields
up to 60 Tesla. To maximize the available experimental space a novel vacuum
tapered seal was developed. The conical plug part of the joint is made out
of G-10, and the matching vacuum can is made out of 1266 Stycast epoxy. The
differential thermal contraction between the parts aids in producing a
superfluid-tight joint. The simple construction of the joint resulted in a
16 mm diameter experimental region.

The main parts of the probe inside the vacuum can are the temperature
regulated block (TRB) and the silicon heat capacity platform with sample,
thin film resistive heater, and bare chip Cernox thermometer. The TRB is
made of 2850 Stycast epoxy chosen for its fair thermal conductivity, and is
thermally connected to the bath via two dozen thin (gauge 44 and 38) 4-inch
long copper wires. The heat capacity platform is suspended with nylon
strings, and electrical connections to thermometer and heater provide a
thermal link as well. The resulting thermal equilibrium time constants for
the TRB and the platform were measured to be on the order of few minutes.
Therefore, during the magnetic field pulse, which lasts for about 2 seconds,
both the TRB and the heat capacity platform can be regarded as thermally
isolated from the bath and each other, and under adiabatic condition. The
third time constant $\tau _{st}$ is that of the heat capacity stage
including sample, platform, thermometer, and heater. The sample's internal
thermal relaxation time constant $\tau _{int}$ can be less than millisecond
at a temperature of a few Kelvin. However it grows rapidly as the
temperature is increased. At the low temperature end $\tau _{st}$ can
increase substantially due to either increase in $\tau _{int}$ (electronic
or nuclear magnetic entropy) or boundary thermal resistance between
different constituents of the stage. The temperature interval between 1 K
and 20 K is therefore a convenient starting point for developing heat
capacity measurements in pulsed magnetic fields.

We use a heat pulse method to measure heat capacity, where a known amount of
heat is delivered to the sample using a chip resistor as a heater element.
The heat capacity stage must come to equilibrium both before and after the
heat pulse is delivered while the magnetic field remains constant. The flat
field plateau of the 60TLP magnet allows this to occur. The temperature of
the stage is measured with a Cernox chip resistance thermometer, which was
calibrated in both DC field up to 30 Tesla and pulsed fields up to 60 Tesla.
The heat capacity of the sample is then determined as the ratio of the heat
delivered to the sample to the change in its temperature. The low ramp rate
of the long-pulse magnet (in comparison with short pulse, capacitively
driven magnet with the total magnet pulse time of about 10 ms) reduces the
unavoidable eddy current heating in metallic samples. However, during
magnetic field sweep the temperature does not stay constant even in the
total absence of eddy current heating, due to the magnetocaloric effect. The
heat capacity stage is thermally isolated from the bath, and remains in
adiabatic condition during the magnetic field pulse. The dependence of the
temperature of the stage on the magnetic field during the pulse is then
given by the expression~\cite{bejan97}

\begin{equation}
\left( {\frac{\delta T}{\delta H}}\right) _S=-{\frac T{C_H}}\left( {\frac{%
\delta M}{\delta T}}\right) _H,  \label{eq:dtdh}
\end{equation}

\noindent where T is temperature, H is magnetic field, M is magnetization,
and C is the specific heat of the sample, and subscripts $S$ and $H$
indicate constant entropy and magnetic field, respectively. Eq.~(\ref
{eq:dtdh}) is used to describe the process of adiabatic demagnization
cooling, where $\left( {\delta M/\delta T}\right) _H$ is negative and
therefore $\left( {\delta T/\delta H}\right) _S$ is positive. Such a system
warms as the field is ramped up, and cools during the ramp down portion of
the magnetic field pulse {\it reversibly}. However, magnetization in general
can also increase with temperature. The sample then would cool during the
ramp up, and warm {\it reversibly} during the ramp down of the magnetic
field. Below we show examples of both types of behavior.

\section{Results}

The first heat capacity experiments in the 60TLP magnet were performed on a
single crystal of metallic YbInCu${\rm _4}$, grown from an In-Cu flux as
described previously.~\cite{sarrao96} This system undergoes a first order
valence transition at 42 K in zero field. The specific volume is increased
by 0.5\% upon cooling through the phase transition,~\cite{felner86} with
accompanying rise in the Kondo temperature $T_K$ from 25 K to 500 K.~\cite
{cornelius97} It is believed that unlike in the case of Ce, where the phase
transition is described within a Kondo-collapse scenario, the valence
transition YbInCu${\rm _4}$ it is driven by the band structure effects.~\cite
{sarrao-ramirez98} The complete magnetic field - temperature phase diagram
was obtained in DC Bitter magnets at NHMFL/Tallahassee and in
capacitor-driven pulsed magnets at NHMFL/Los Alamos.~\cite{immer97} This
work showed that the transition can be suppressed down to T = 0 K with an
applied field of 34.3 Tesla.

The length of the flat top can be close to 0.5 s in the 60TLP magnet for
magnetic fields less than or equal to 35 Tesla. When $\tau _{st}$ is much
smaller than the length of the plateau, a sequence of heat pulses can be
delivered to it within the flat portion of the field profile, with
sufficient time for the calorimeter to come to equilibrium before and after
each of the heat pulses. This situation is illustrated by Fig.~\ref{train}%
(a) for a field of 20 Tesla, where a sequence of five 10 ms long heat pulses
were delivered to the heat capacity stage during the plateau of a single
magnetic field pulse. The thermometer comes to equilibrium after the heat
pulse is delivered well before the next heat pulse, and temperature is
determined before and after each of the pulses. In this way a series of five
C$_H$(T) data points is collected in a single ''shot'' experiment, as the
initial temperature for each of the heat capacity experiments is increased
due to the previous heat pulse. The data from this and one zero field
''shot'' is shown in Fig.~\ref{train}(b). We fit the data with a sum of
T-linear (electronic) and T-cubic (phononic) terms $AT+BT^3$. For zero field
we obtain $A=49.5\pm 0.4$ ${\rm mJ/molK^2}$ and $B=(0.85\pm 0.03)$ ${\rm %
mJ/molK^4}$. The value of A is in excellent agreement with available data in
the literature.~\cite{sarrao-ramirez98} At 20 Tesla we obtain $A=80\pm 5$ $%
{\rm mJ/molK^2}$ and $B=(0.81\pm 0.07)$ ${\rm mJ/molK^4}$. The magnitude of
the cubic term due to phonons is field-independent as expected. The increase
of the linear term with field is likely due to scaling with magnetic field
observed for various properties of YbInCu$_4$.~\cite{immer97,jaime99}

Another way to increase the data acquisition rate relies on the programmable
nature of the 60TLP magnet. A series of plateaus at different magnetic
fields can be produced within a single experiment. Fig.~\ref{multi} displays
the data for one such experiment on YbInCu$_4$, with four plateaus at 25,
30, 35, and 40 Tesla, each 130 ms long. At each of the magnetic field
plateaus the heat pulse is applied to the stage, and heat capacity
experiment is performed. In addition, as the field was changed between 30
and 35 Tesla through the first order phase boundary, the temperature of the
sample was observed to go down on the up-sweep, and up on the down-sweep,
due to the magnetocaloric effect. These features are very sharp, and allow
direct determination of the phase diagram of YbInCu$_4$. This is yet another
complementary way to collect data using the heat capacity apparatus. It was
not possible to measure heat capacity in the high field phase due to a large
increase in $\tau _{st}$.

Preliminary measurements of heat capacity of UBe$_{13}$ and Ce$_3$Bi$_4$Pt$%
_3 $ were performed up to 60 Tesla. Fig.~\ref{mc2} shows temperature
variation of UBe$_{13}$ and Ce$_3$Bi$_4$Pt$_3$ as a function of field during
magnetic field pulses to 60 T. This figure illustrates the magnetocaloric
effect under adiabatic conditions of our apparatus. The temperature of the
UBe$_{13} $ sample increases from 4 K up to 10 K during the ramp up, and
returnes to 4 K during the ramp down in a very reversible fashion. The
opposite is true for Ce$_3$Bi$_4$Pt$_3$, which is colder at 60T than at 0 T.
With available magnetization data it should be possible to calculate the
specific heat of these compounds during the ramp portions of the field
pulse, providing information in addition to the direct heat capacity
measurements at field plateaus.

\section*{Conclusion}

We have demonstrated the feasibility of heat capacity measurements in the
pulsed magnetic fields provided with the 60 Tesla Long Pulse magnet at
NHMFL/LANL. Direct measurement of heat capacity at field plateaus was
clearly demonstrated for a variety of compounds. It appears that thermal
equilibrium can be achieved in some compounds even during the field sweep,
given the low ramp rates which the 60TLP magnet is capable of achieving. In
this situation specific heat data can be obtained via the magnetocaloric
effect from the temperature vs. field traces and magnetization data for such
compounds. Other types of thermal relaxation-related experiments like
thermal conductivity and Seebeck effect are also under development.

\section*{Acknowledgments}

This work was conducted under the auspices of the Department of Energy. It
was also supported by the In-House Research Program of the NHMFL.

\begin{figure}[tbp]
\caption{Multiple heat capacity experiments on YbInCu$_4$ during a single 20
Tesla 0.5 s long pulse. a) Dashed line - magnetic field. Solid line -
voltage applied to the resistive heater. $\bullet$- Thermometer's
temperature. b) $\Box$ - specific heat at H = 0 T collected with our probe.
Filled squares - specific heat at 20 T obtained from data in a).}
\label{train}
\end{figure}

\begin{figure}[tbp]
\caption{Staircase pulse shape for specific heat measurement of YbInCu$_4$.
Dashed line - magnetic field. Solid line - Voltage applied to the resistive
heater. $\bullet $ - Thermometer's temperature.}
\label{multi}
\end{figure}

\begin{figure}[tbp]
\caption{Tempertature vs. field for UBe$_{13}$ (solid lines) and Ce$_3$Bi$_4$%
Pt$_3$ (dashed lines) during field ramp up to 60 Tesla and back to 0. Heat
pulses were applied at the 60 T plateaus for Ce$_3$Bi$_4$Pt$_3$.}
\label{mc2}
\end{figure}

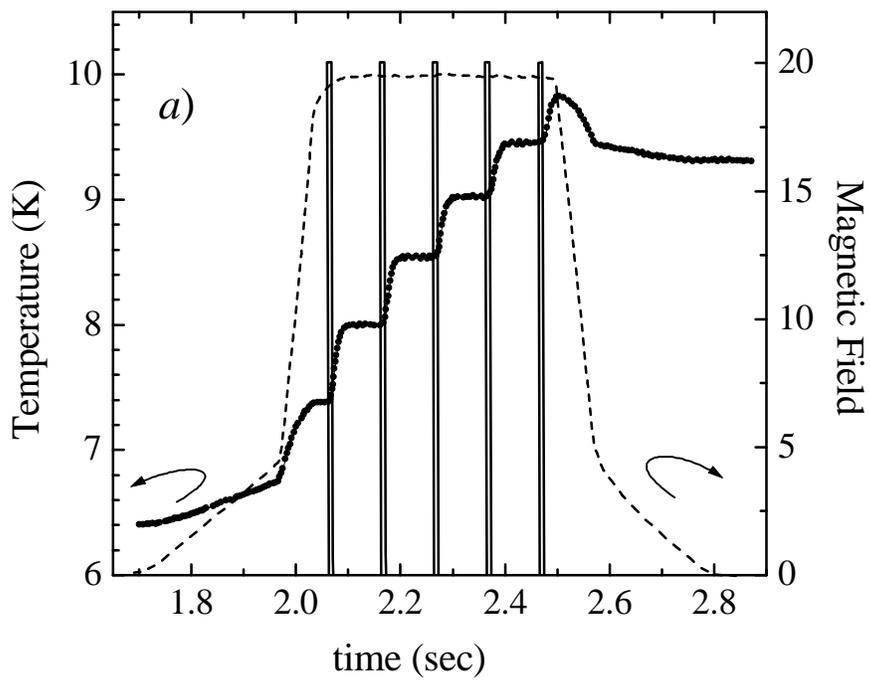
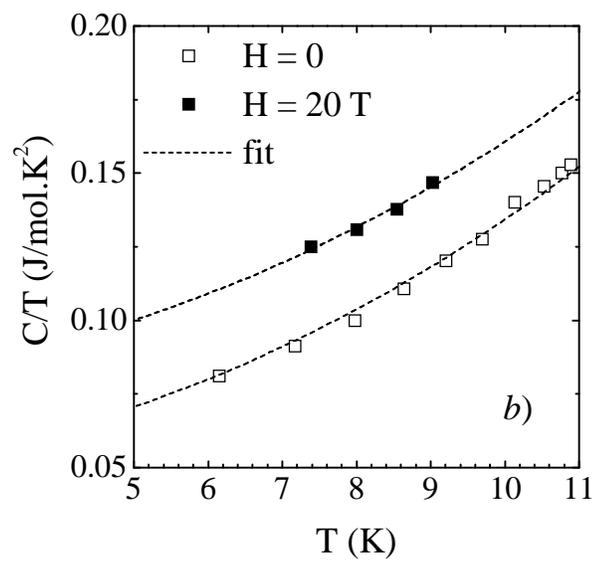
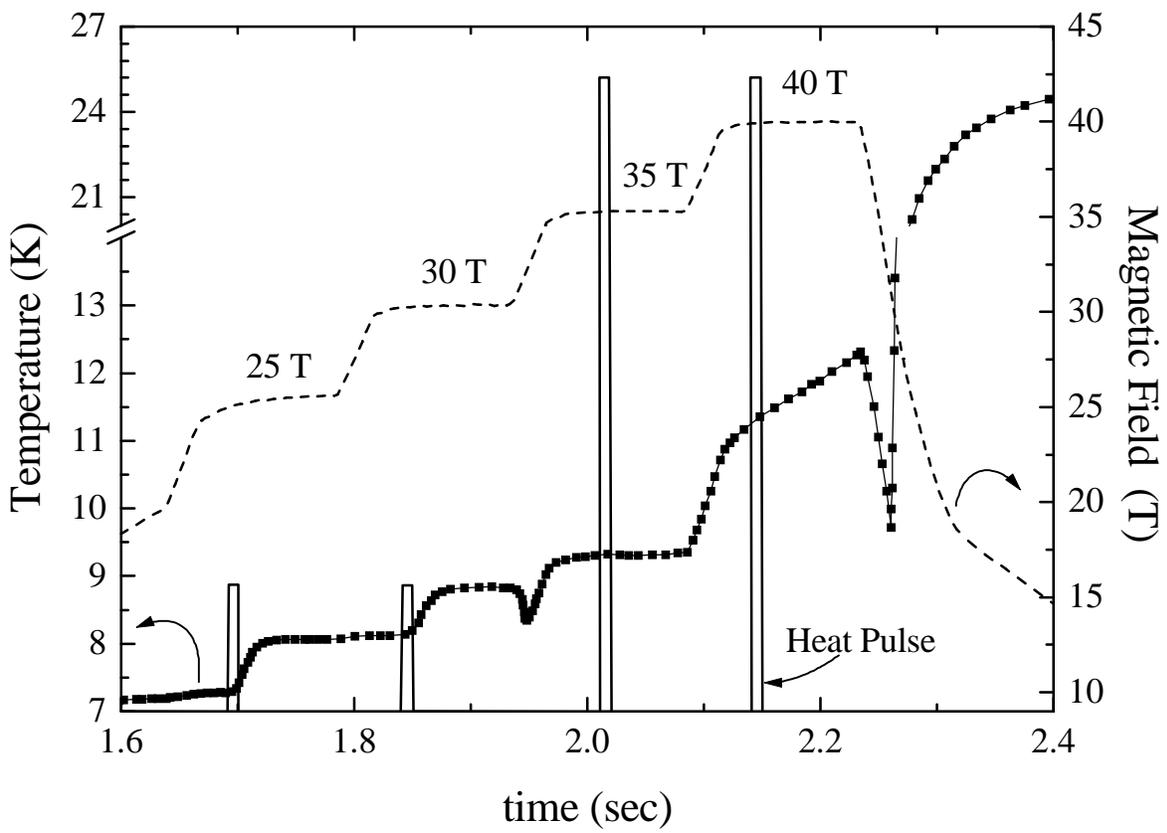
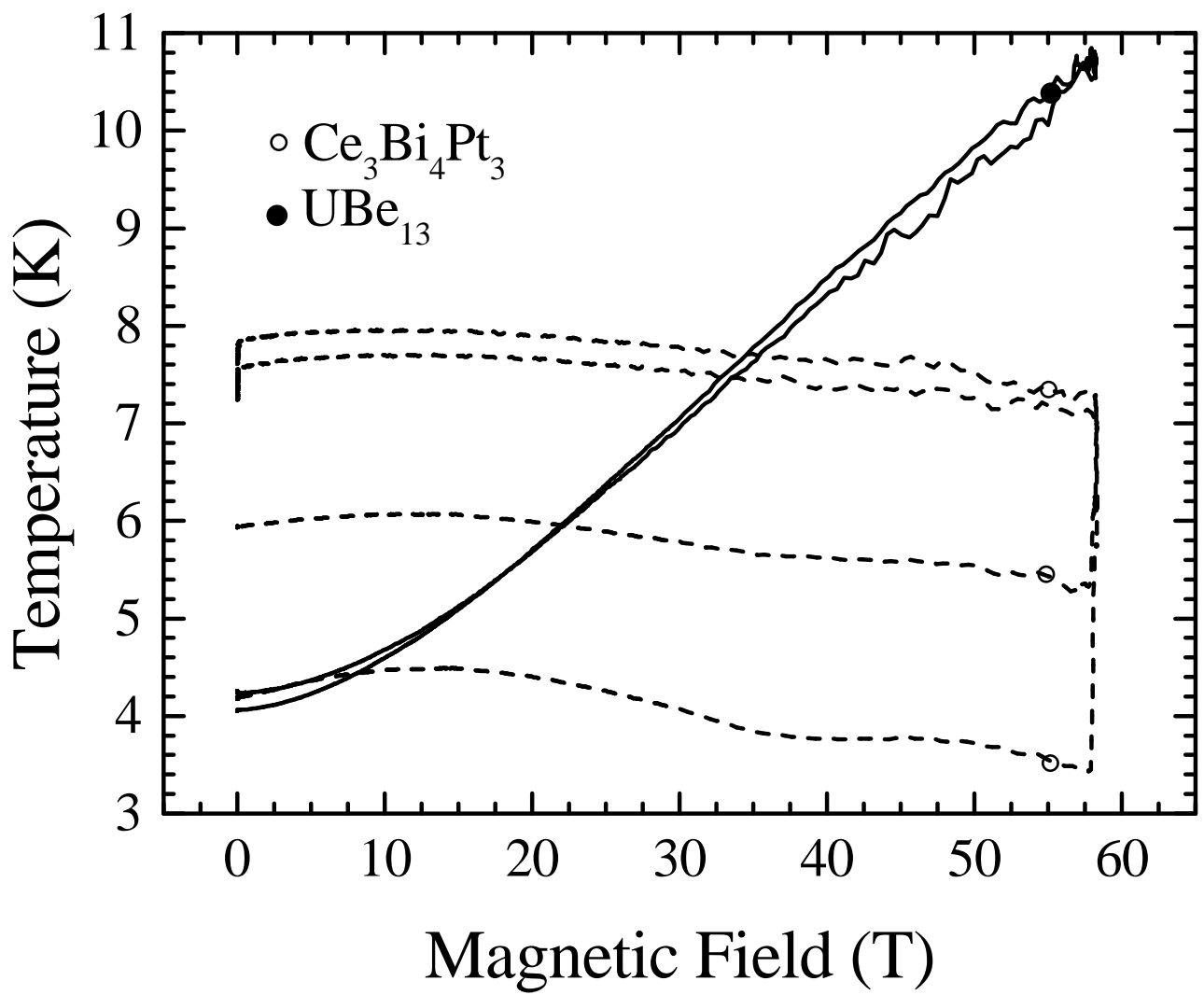
\end{document}